\documentclass[MSNbibl,nameyear,dvips]{arxstspdf}
\usepackage{flushend}
\usepackage{stfloats}
\usepackage{graphicx,url,breakurl}
\volume{29}
\issue{1}
\pubyear{2014}
\firstpage{58}
\lastpage{68}
\doi{10.1214/13-STS419} 

\makeatletter

\newcommand{\rright}{\right}
\newcommand{\lleft}{\left}

\newcommand{\btheta}{\bolds{\theta}}
\newcommand{\bdelta}{\bolds{\delta}}
\newcommand{\bphi}{\bolds{\phi}}

\makeatother

\begin{document}
\begin{frontmatter}

\title{Bayesian Population Projections for the United
Nations}
\runtitle{Bayesian Population Projections}

\begin{aug}
\author[a]{\fnms{Adrian E.} \snm{Raftery}\corref{}\ead[label=e1]{raftery@uw.edu}\ead[url,label=u1]{www.stat.washington.edu/raftery}},
\author[b]{\fnms{Leontine} \snm{Alkema}\ead[label=e2]{alkema@nus.edu.sg}}
\and
\author[c]{\fnms{Patrick} \snm{Gerland}\ead[label=e3]{gerland@un.org}}
\runauthor{A. E. Raftery, L. Alkema and P. Gerland}

\affiliation{University of Washington, National University of Singapore
and United Nations}

\address[a]{Adrian E. Raftery is Professor of Statistics and Sociology, Department of Statistics,
University of Washington,
Box 354322, Seattle, Washington 98195-4322, USA \printead{e1}.}
\address[b]{Leontine Alkema is Assistant Professor, Department of Statistics and
Applied Probability and Saw Swee Hock School of Public Health,
National University of Singapore, Singapore 117546 \printead{e2}.}
\address[c]{Patrick Gerland is Population Affairs Officer, Population Estimates and
Projections Section, United
Nations Population Division, New York, New York 10017, USA
\printead{e3}.}

\end{aug}

%
\begin{abstract}
The United Nations regularly publishes projections of the populations
of all the world's countries broken down by age and sex. These
projections are the de facto standard and are widely used by
international organizations, governments and researchers. Like almost
all other population projections, they are produced using the standard
deterministic cohort-component projection method and do not yield
statements of uncertainty. We describe a Bayesian method for producing
probabilistic population projections for most countries which are
projections that the United Nations could use. It has at its core
Bayesian hierarchical models for the total fertility rate and life
expectancy at birth. We illustrate the method and show how it can be
extended to address concerns about the UN's current assumptions about
the long-term distribution of fertility. The method is implemented in
the R~packages \texttt{bayesTFR}, \texttt{bayesLife}, \texttt{bayesPop}
and \texttt{bayesDem}.
\end{abstract}

%

\end{frontmatter}

\section{Introduction}
The United Nations (UN) publishes projections of the populations
of all countries broken down by age and sex, updated every
two years in a publication called the \textit{World Population Prospects} (WPP).
It is the only organization to do so. These projections are used
by researchers, international organizations and governments,
particularly with less developed statistical systems,
and researchers. They are used for planning, social and health research,
monitoring development goals, and as inputs to other forecasting models
such as those used for predicting climate change and its impacts
(\cite{IPCC2007}; \cite{Seto2012}).
They are the de facto standard (\cite{LutzSamir2010}).

Like almost all other population projections,
the UN's projections are produced using the standard
cohort-component projection method
(\citeauthor{Whelpton1936},\linebreak[4]  \citeyear{Whelpton1936}; Leslie, \citeyear{Leslie1945}; \cite{Preston2001}).
This is a deterministic method based on an age-structured version
of the basic demographic identity that the number of people
in a country at time $t+1$ is equal to the number at time $t$
plus the number of births, minus the number of deaths, plus the
number of immigrants, minus the number of emigrants.

The UN projections are based on assumptions about future fertility,
mortality and
international migration rates; given these rates, the UN produces
the ``Medium'' projection, a single
value of each future population number with no statement of uncertainty.
The UN also produces ``Low'' and ``High'' projections using total fertility
rates (the average number of children per woman) that are, respectively,
half a child lower and half a child higher than the
Medium projections. These are alternative scenarios that also have
no probabilistic interpretation.

Scientists, including researchers working on climate change, have long
expressed interest in UN population projections that would include
statistical uncertainty intervals. This was first expressed in 1986 by
a call to incorporate a probabilistic element in UN projections and to
probabilistically specify the range of error (\cite{ElBadryKono1986}).
Independent evaluations of UN projections (\cite{NRC2000}; \cite{Keilman2002})
and expert-based probabilistic projections for the world and major
regions
(Lutz, Sanderson and Scherbov,
\citeyear{Lutzetal98,Lutzetal04book,Lutzetal07projDB}) have
further highlighted the desirability of uncertainty bounds.

Responding to the call for the inclusion of uncertainty in populations
projections, the UN is interested in producing probabilistic population
projections
for all countries; here we describe the current state of an ongoing
effort to develop a methodology for doing so.
Our method builds on previous work on time series methods for
probabilistic population projections (\cite{NRC2000}),
particularly the work of Ronald D. Lee and his collaborators
(\cite{LeeCarter1992}; \cite{LeeTuljapurkar1994}; \cite{Lee2011}).

In Section \ref{sect-current} we summarize the current UN approach and
in Section \ref{sect-bayes} we describe our probabilistic approach.
In Section \ref{sect-newtfr} we consider how a modification to the method
could accommodate disagreement about the long-term behavior of fertility
assumed in the model, and in Section \ref{sect-discussion} we
discuss the contribution of Bayesian thinking to the method.

\section{Current UN Population Projection Methodology}
\label{sect-current}
We now outline the UN's current (deterministic) population projection method,
as used in the \textit{World Population Prospects 2008}
(\cite{UN2009}) and described by \citet{UN2006}.
The most recent UN projections published in the
\textit{World Population Prospects 2010} (\cite{UN2011})
incorporate some aspects of the new methods we will describe here.
Thus, we will refer to the 2008 WPP method as the ``current'' method.

\subsection{Cohort Component Projection Method}
At the heart of the UN's current population projection method
lies the cohort component projection or Leslie matrix method.
To fix ideas, we describe a simplified version here.
We consider one sex (female) and divide the population into $N$
$k$-year age groups; those in the $x$th age group are aged from
$k(x-1)$ years to $(kx-1)$ years.
The projection is done by $k$-year time periods,
where $k$ is typically 5 or 1 (in our work we use $k=5$),
and the beginning of the $t$th time
period will be referred to as time $t$.

We let $n_{x,t}$ be the number of
females in the $x$th age group at time $t$.
We let $S_{x,t}$ be the survival ratio for the $x$th age group
in the $t$th period, that is, the proportion of the females in the
$x$th age group at time $t$ who are still alive at time $t+1$.
We let $B_{x,t}$ be the
number of female offspring of females in the $x$th age group at time
$t$ who
are born in the $t$th period and survive to time $t+1$, divided by $n_{x,t}$.
Finally, we denote net migration by $m_{x,t}$, equal to the number of
immigrants during the $t$th period who were in the $x$th age group
at time $t$ and are still in the population at time $t+1$,
minus the corresponding number of emigrants.
For the highest age group, $N$, we assume that those who
survive stay in the same age group at the next time point.

Then the model is simply a series of deterministic accounting identities:
$n_{1,t+1} = \sum_{x=1}^N B_{x,t} n_{x,t} + m_{1,t}$,
$n_{x+1,t+1} = S_{x,t} n_{x,t} + m_{x+1,t}$ for $x=1,\ldots,N-2$,
and $n_{N,t+1} = S_{N-1,t}n_{N-1,t} + S_{N,t} n_{N,t} + m_{N,t}$.
If we define the projection matrix for period $t$, $\mathbf{P}_t$, by
\[
\mathbf{P}_t = \lleft[ \matrix{ B_{1,t} & B_{2,t} &
\cdots& B_{N-1,t} & B_{N,t}
\cr
S_{1,t} & 0 & \cdots& 0
& 0
\cr
0 & S_{2,t} & \cdots& 0 & 0
\cr
\vdots& \vdots& \ddots&
\vdots& \vdots
\cr
0 & 0 & \cdots& S_{N-1,t} & S_{N,t}} \rright],
\]
then the model can be rewritten in matrix form as
%
\begin{equation}\label{eqn-leslie}
\mathbf{n}_{t+1} = \mathbf{P}_t \mathbf{n}_t +
\mathbf{m}_t,
\end{equation}
where $\mathbf{n}_t = ( n_{1,t},\ldots, n_{N,t} )^T$ and
$\mathbf{m}_t = ( m_{1,t},\ldots,\allowbreak m_{N,t} )^T$.
This can be applied recursively to obtain population projections.
It can be extended in a fairly straightforward way to project
two-sex populations.

The formulation (\ref{eqn-leslie}) is due to \citet{Leslie1945}.
The deterministic analysis of (\ref{eqn-leslie})
and its use for population projections are
the subject of classical mathematical or formal demography; see, for example,
\citet{Preston2001}, \citet{KeyfitzCaswell2005} and
\citet{Caswell2006}.

The use of (\ref{eqn-leslie}) for population projections requires
that values of future age-specific mortality, fertility and migration
be specified for each future time period to be projected.
This is the hard part, and most of the uncertainty about future
population is due to uncertainty about these future quantities.

\begin{figure}[b]

\includegraphics{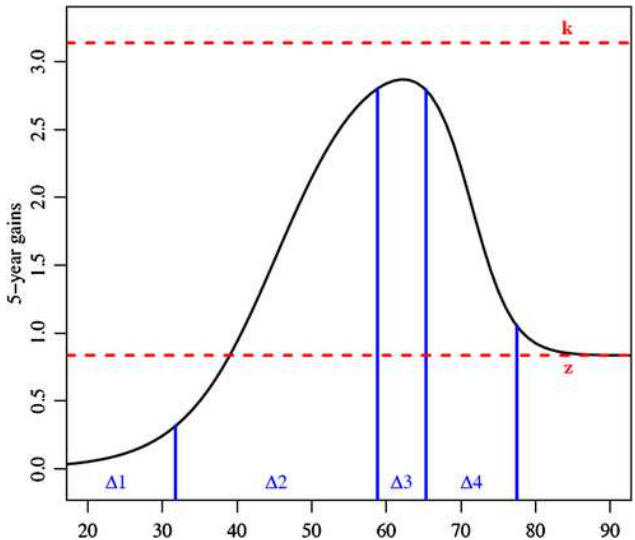}

\caption{Example of a double logistic function used for projecting the five-year
gains in life expectancy, plotted against life expectancy.}
\label{fig-Fig1}
\end{figure}

\subsection{Projecting Mortality and Fertility Rates}
The UN's current method generates assumptions about future age-specific
fertility and mortality rates for most countries by projecting forward
the overall level of future fertility or mortality, and then converting the
overall levels to age-specific rates.

The UN's current method for projecting life expectancy at birth
(hereafter just referred to as life expectancy) for most
countries is as follows. Five-year gains in life expectancy
for country $c$ in time period $t$, $\ell_{c,t}$,
are projected using a deterministic
double logistic function, namely,
%
\begin{equation}\label{eq-mortalityDL}
\ell_{c,t+1} = \ell_{c,t} + g\bigl(\ell_{c,t}|
\btheta^{c}\bigr),
\end{equation}
where five-year gains $g(\ell_{c,t}| \btheta^{c})$ are given by
%
\begin{eqnarray}\label{eq-g}
&&
g\bigl(\ell_{c,t}| \btheta^{c}\bigr)\hspace*{-7pt} \nonumber\\
&&
{\fontsize{10.6pt}{12pt}\selectfont{\mbox{$\displaystyle \quad= \frac{k^{c}}{1+\exp(-
{A_1}(\ell_{ct}-\Delta_{1}^{c} - A_2
\Delta_{2}^{c})/{\Delta_{2}^{c}})}$}}}\hspace*{-7pt}\\
&&
{\fontsize{10.6pt}{12pt}\selectfont{\mbox{$\displaystyle\qquad{} + \frac{z^{c}-k^{c}}{1+\exp(-{A_1}
(\ell_{ct}-\sum_{i=1}^3\Delta_i^{c}-A_2 \Delta_{4}^{c})/{\Delta_{4}^{c}})}$}}}.
\hspace*{-7pt}\nonumber
\end{eqnarray}
In (\ref{eq-g}),
$\btheta^{c} = (\Delta_{1}^{c},\Delta_{2}^{c},\Delta_{3}^{c},\Delta
_{4}^{c},k^{c},z^{c})$ are the six parameters of the double logistic function
for country $c$, whose meaning is illustrated in Figure
\ref{fig-Fig1},\vadjust{\goodbreak}
and $A_1$ and $A_2$ are constants. The parameters to be used for a
given country are chosen by the UN analyst for that country from a list
of five predetermined patterns that represent different rates of
improvement in life expectancy.

The most used measure of the overall level of fertility is the
total fertility rate (TFR) for country $c$ at time~$t$,
defined as $f_{c,t} = k\sum_{x=1}^{N-1} F_{c,x,t}$, where
$F_{c,x,t}$ is the fertility rate in country $c$ for age group $x$ at
time $t$.
The TFR is the average number of children a woman would bear in her
life if exposed to the age-specific fertility rates prevalent at time $t$.

The UN's current method for projecting TFR takes account of
several empirical regularities.
The past century has been dominated by the fertility transition,
a~shift from high fertility and high mortality to low fertility and
low mortality, that started in Europe and North America in the late 19th
century and in East Asia in the mid 20th century. It has now started in
almost all countries and is complete in many (\cite{Hirschman1994}).
The patterns of change in different countries have been similar.
The TFR starts from a high level that differs among countries but
is typically between 4 and 8, and then starts to decline
slowly. The pace of decline reaches a peak about half way through
the transition.
Then the pace of decline slows, stopping some time after the TFR goes
below the
replacement level of about 2.1 children per woman.
In several low-fertility countries, a slow increase has been observed
after this point.

The UN has projected five-year decrements in the TFR using a double
logistic function, similar to the function used for projecting gains in
life expectancy.
The parameters to be used for a given country are chosen by the UN
analyst for that country from a list
of three predetermined patterns. In the projections, the TFR is held
constant once it reaches 1.85 children, which represents the
deterministic ultimate fertility level or asymptote. For countries
where the TFR is below 1.85 at the start of the projection, the TFR is
projected to increase by 0.05 children per 5-year period until the
ultimate level is reached.

Projected values of the total fertility rate and life
expectancy are converted to age-specific fertility and mortality rates
based on past patterns or model life tables, and population projections
are produced using the cohort component method.
Finally, high and low variants are produced by increasing or decreasing
the total fertility rate in each future period by half a child.

\section{Bayesian Population Projections}
\label{sect-bayes}
The UN's current projection method does not yield an assessment
of uncertainty about future population quantities. It is somewhat
subjective because the double logistic functions used have been selected
by the analyst from a small number of predetermined possibilities
rather than estimated from the data. It is also somewhat rigid
in that the set of double logistic functions used is small and may
not cover a full range of realistic future possibilities.

To address these issues, we have developed a Bayes\-ian probabilistic
population projection method. This involves building Bayesian hierarchical
models to project the total fertility rate and life expectancy,
each of which produces a large number of possible future trajectories
from the posterior predictive distribution. These are then input
to the cohort component projection method to provide a posterior
predictive distribution of any future population quantity of interest.
We now briefly describe the method.

\begin{figure}[b]

\includegraphics{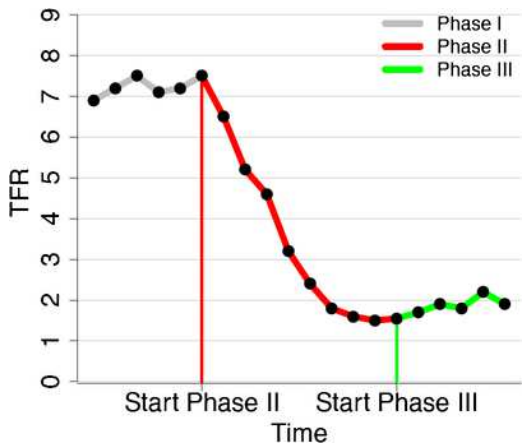}

\caption{The three phases of the model for total
fertility---Phase~I: pre-demographic transition; Phase II: fertility
transition; Phase III; post-transition recovery.} \label{fig-Fig2}
\end{figure}

\subsection{Bayesian Fertility Projection Model}
We model the typical evolution of a country's fertility over time as consisting
of three phases, shown in Figure \ref{fig-Fig2} (\cite{Alkema2011}).
Phase I precedes the beginning of the fertility transition
and is characterized by high fertility that is stable or increasing.
All countries have now completed this phase, and so it is not of
interest for projections; we do not model it further.
Phase II consists of the fertility transition during which
fertility declines from high levels to below the replacement level of 2.1
children per woman. Phase III is the post-fertility transition period.

To model fertility declines in Phase II, we use a double logistic function,
but with some modifications. First, to make the model stochastic,
we add a heteroscedastic error term. Second, we allow the parameters
$\theta^{(c)}$ to vary continuously rather than being restricted to
a
small number of possibilities. Third, we model the values of a
parameter for
different countries as arising from a ``world'' distribution.
This leads to estimates that borrow strength from data for other countries
and makes the model hierarchical. This is important because, for a single
country, the data are sparse (at most 12 five-year periods for most
countries), and estimation
of the country-specific double-logistic curve can be unstable,
as it involves estimating five parameters from 12 or fewer data points.
The hierarchical model stabilizes the estimation.

The resulting model is as follows:
%
\begin{equation}\label{eq-fct}
f_{c,t+1} = f_{c,t} - r\bigl(f_{c,t}|
\bdelta^c\bigr) + a_{c,t},
\end{equation}
where the five-year decrement $r(f_{c,t}|\bdelta^c)$ is given by
%
\begin{eqnarray}\label{eq-fertilityDL}
&&
r\bigl(f_{c,t}|\bdelta^c\bigr) \hspace*{-6pt}\nonumber\\
&&{\fontsize{10.4pt}{12pt}\selectfont{\mbox{$\displaystyle
\quad= \frac{-d^{c}}{1+\exp(-{2\ln
(9)}(f_{c,t}-
\sum_{i=2}^4\bigtriangledown_{i}^{c}
+ 0.5\bigtriangledown_{1}^{c})/{\bigtriangledown_{1}^{c}} )}$}}}\hspace*{-6pt} \\
&&{\fontsize{10.4pt}{12pt}\selectfont{\mbox{$\displaystyle\qquad{}+
\frac{d^{c}}{1+\exp(-{2\ln(9)}(f_{c,t}-\bigtriangledown_{4}^{c} - 0.5\bigtriangledown
_{3}^{c})/{\bigtriangledown
_{3}^{c}} )}$}}}\hspace*{-6pt}\nonumber
\end{eqnarray}
with $\bdelta^{c} =
(\bigtriangledown_{1}^c,\bigtriangledown_{2}^c,\triangle
_{3}^c,\bigtriangledown_{4}^c,d^c)$ being a vector of country-specific
parameters and $a_{c,t} \stackrel{\mathrm{ind}}{\sim}
N(0,\sigma(t,\break  f_{c,t})^2)$, where $\sigma(t,f_{c,t})$ is a function that
describes how the error standard deviation changes with fertility level
and time period.

The country-specific parameters, $\bdelta_c$, are assumed
to be drawn from a world distribution whose parameters (or hyperparameters)
themselves have a diffuse prior distribution, namely,
$\bdelta_c \stackrel{\mathrm{i.i.d.}}{\sim} h(\cdot,\bphi)$, where
$\bphi \sim p(\bphi)$.
The resulting Bayesian hierarchical model is estimated using Markov chain
Monte Carlo.

We define a country as having entered Phase III once two consecutive
five-year increases below a TFR of 2 children have occurred.
By this definition 21 countries had entered Phase III by 2010:
19 European countries, the USA and Singapore.
For these countries, TFR has tended to increase back toward
replacement level after they entered Phase II, reversing the secular
trend of fertility decline. This is by now a well-documented trend
(\cite{Myrskyla2009}).

To model this, we used a single first-order autoregressive model
with long-term mean $\mu$ equal to the approximate replacement fertility
level of 2.1 for all countries in Phase III, namely,
%
\begin{equation}\label{eq-PhaseIII}
f_{c,t+1} - \mu= \rho(f_{c,t} - \mu) + b_{c,t},
\end{equation}
where $b_{c,t} \stackrel{\mathrm{i.i.d.}}{\sim} N(0,\sigma_b^2)$.
The parameters $\rho$ and $\sigma_b$ were estimated by maximum likelihood
from the 54 time periods observed in the 21 countries
that have entered Phase III, yielding $\hat{\rho} = 0.89$ and
$\hat{\sigma}_b = 0.10$. The estimated value of $\rho$ gives expected
increases that are similar to the current UN increments of 0.05 children
for each five-year period until the TFR equals 1.85.

\subsection{Bayesian Life Expectancy Projection Model}
We model female life expectancy similarly to Phase~II total fertility.
We use the UN's double logistic function to project expected gains, but
we add a heteroscedastic error term, we allow the parameters of the
country-specific double logistic functions
to vary continuously among countries rather than being restricted
to five pre-assigned possibilities, and we assume that the
double logistic parameters are draws from a common ``world'' distribution
(\cite{RafteryChunn2012}).

The resulting Bayesian model is as follows:
%
\begin{equation}\label{eq-lctstoch}
\ell_{c,t+1} = \ell_{c,t} + g\bigl(\ell_{c,t}|
\btheta^{(c)} \bigr) + e_{c,t},
\end{equation}
where
%
\begin{eqnarray}\label{eq-gBayes}
&&
g\bigl(\ell_{c,t}| \btheta^{c}\bigr)\hspace*{-7pt}\nonumber\\
&&{\fontsize{10.6pt}{12pt}\selectfont{\mbox{$\displaystyle
\quad = \frac{k^{c}}{1+\exp(-
{A_1}(\ell_{ct}-\Delta_{1}^{c} - A_2 \Delta_{2}^{c})/{\Delta_{2}^{c}})}$}}} \hspace*{-7pt}\\
&&
{\fontsize{10.6pt}{12pt}\selectfont{\mbox{$\displaystyle \qquad{}+
\frac{z^{c}-k^{c}}{1+\exp(-{A_1}
(\ell_{ct}-\sum_{i=1}^3\Delta_i^{c}-A_2 \Delta_{4}^{c})/{\Delta_{4}^{c}})}$}}}.\hspace*{-7pt}\nonumber
\end{eqnarray}
In (\ref{eq-gBayes}),
$\btheta^{c} = (\Delta_{1}^{c},\Delta_{2}^{c},\Delta_{3}^{c},\Delta
_{4}^{c},k^{c},z^{c})$,
$e_{c,t} \stackrel{\mathrm{ind}}{\sim} N(0,\break  \omega(\ell_{c,t})^2)$,
where $\omega(\cdot)$ is a smooth function representing how the error
standard deviation depends on the current level of life expectancy,
and $A_1$ and $A_2$ are constants.
The country-specific parameters are assumed to be drawn from
world distributions, as follows:
%
\begin{eqnarray}
\label{eq-Deltaic}
\quad\Delta_{i}^{c} &\stackrel{\mathrm{i.i.d.}} {\sim}&
\mathrm{TN}_{[0,100]}\bigl(\Delta_{i}, \sigma_{\Delta_i}^2
\bigr),\quad i=1,\ldots,4,
\\
\label{eq-kc}
\quad k^{c} & \stackrel{\mathrm{i.i.d.}} {\sim} & \mathrm{TN}_{[0,10]}\bigl(k,
\sigma_{k}^2\bigr),
\\
\label{eq-zc}
\quad z^{c} & \stackrel{\mathrm{i.i.d.}} {\sim} & \mathrm{TN}_{[0,1.15]}\bigl(z,
\sigma_{z}^2\bigr),
\end{eqnarray}
where TN$_{[a,b]} (\mu, \sigma^2)$ denotes a truncated normal distribution
with mean parameter $\mu$ and standard deviation parameter $\sigma$,
truncated to lie between $a$ and~$b$.

The world hyperparameters on the right-hand sides of equations
(\ref{eq-Deltaic})--(\ref{eq-zc}) are given diffuse prior distributions,
with one notable exception, namely, $z$.
The country-specific parameter $z^c$
is the asymptotic linear increase of life expectancy in country $c$
per five-year period, and this is restricted to be less than 1.15,
which is highly informative.
This is based on the empirical fact that over the past 170 years the maximum
country-specific life expectancy in year $y$ has been increasing highly
linearly with $y$ (\cite{OeppenVaupel2002}); 1.15 years per five-year period
is the upper bound of a 99.9\% confidence interval for the rate of
increase. Accordingly, the prior distribution of $z$ is also bounded
above by 1.15.

Male life expectancy is highly correlated with female life expectancy
and is almost invariably lower. We therefore project
female and male life expectancy jointly, by first projecting female
life expectancy using the Bayesian hierarchical model described above
and then projecting the gap between them. On average, the gap tends
to increase as a function of female life expectancy as female life
expectancy increases
up to about 75 years, and then tends to decrease. It also has extreme values,
often corresponding to conflicts when male life expectancy is more affected
than female. We represent this using the regression model of
\citet{Lalic2011} with $t$-distributed errors for $G_{c,t}$, the
gap in
country $c$ at time period $t$.

The model is as follows:
\[
G_{c,t+1} = \min\bigl\{ \bigl(G^*_{c,t+1}\bigr)_+, 18 \bigr\},
\]
where
\begin{eqnarray*}
G^*_{c,t+1} &=& \cases{ \beta_{0}+\beta_{1}
\ell_{c, 1950-1955}\cr
\quad{}+\beta_{2}G_{c,t}+ \beta_{3}
\ell_{c,t}\cr
\quad{}+ \beta_{4}(\ell_{c,t} - 75)_+ \cr
\quad{}+
\epsilon_{c,t}, & if $\ell_{c,t} \leq M$,
\cr
\gamma_{1}G_{c,t}+\epsilon_{c,t}, & if $
\ell_{c,t} > M$,}
\\
\epsilon_{c,t} &\stackrel{\mathrm{i.i.d.}} {\sim}& t\bigl(\mu=0,
\sigma^{2}=0.0665,\nu=2\bigr)
\end{eqnarray*}
and $M=86.2$ years, the highest life expectancy recorded in the WPP 2010.
The gap is restricted to be no more than 18 years, which is slightly
above the highest value observed to date. The model was estimated
by maximum likelihood.

\subsection{Bayesian Population Projections}
The methods just described formed the basis for probabilistic population
projections for 159 countries, comprising just under 90\% of the
world's population in 2010. The 38 countries with generalized
HIV/AIDS epidemics were not included because they have very different
mortality patterns and require special treatment. Thirty small countries
or areas with populations under 100,000 were also excluded.

\begin{figure*}

\includegraphics{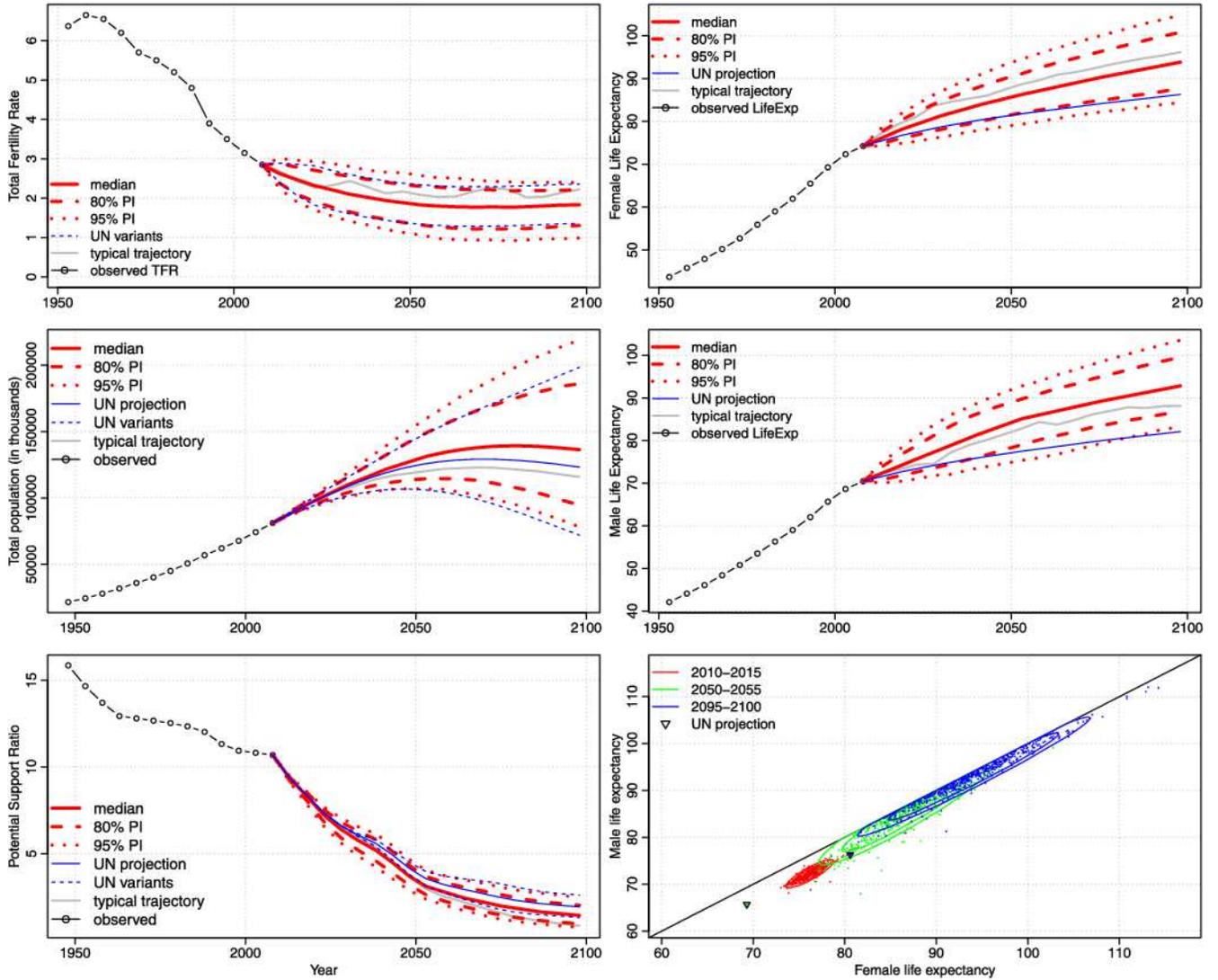}

\caption{Bayesian probabilistic population projections for Egypt, 2010--2100:
major population indicators.
Left, top to bottom: total fertility rate; total population;
potential support ratio (20--64 population$/$65$+$ population).
Right, top to bottom: female life expectancy; male life expectancy;
joint predictive distribution of female and male life expectancy for
2010--2015, 2050--2055 and 2095--2100.
The Bayesian predictive distributions are shown in red: median---solid;
80\% projection interval---dashed; 95\% projection interval---dotted.
The UN WPP 2010 medium projection is shown as a solid blue line
and the UN WPP 2010 high and low projections are shown
as dashed blue lines. The solid grey line represents a typical trajectory.}
\label{fig-Egypt}
\end{figure*}

To produce the probabilistic projections,
2000 trajectories of the total fertility rate for
each five-year period from 2010 to 2100,
and 2000 joint trajectories of female and male life expectancy were simulated
from their posterior predictive distributions. These were then
converted to
age-specific fertility and age- and sex-specific mortality rates using
established UN methods, and input to the cohort-component method.
Current UN assumptions about future international migration were used.
This yielded joint probabilistic projections of any future population
quantity of interest.

The method was described in more detail by \citet{Raftery2012},
which also reported the results of a study assessing the model
by estimating it from data for 1950--1990 and using it to project
population for all 159 countries in the 20-year period 1990--2010.
The projections of total population,
total fertility rate, and female and male life expectancy were
reasonably accurate and the projection intervals were reasonably
well calibrated.

The results for Egypt are shown in Figure \ref{fig-Egypt}.
For the total fertility rate, the UN high and low variants turn out
to be similar to the limits of the Bayesian pointwise 80\% projection intervals.
For life expectancy, the current UN projections do not provide any
assessment of uncertainty or even scenarios.
The Bayesian approach suggests higher future life expectancy,
but the current UN projection is within the Bayesian 95\% interval
for most years. The median Bayesian projection of total population
in 2100 is about 10\% higher than the UN's WPP 2010 projection,
but this has to be seen in the context of the considerable uncertainty
about Egypt's total population at century's end.
The Bayesian 80\% interval for Egypt's population in 2100 ranges from
96 to 184 million.

Perhaps the most striking result is the projected
trend in the potential support ratio, equal to the number of people
aged 20--64 per person aged 65 or over. This\vadjust{\goodbreak} can be roughly interpreted
as the number of workers per retiree and is important, for example,
for planning old-age social security systems.
In Egypt this is currently 10.7, but is projected to decline dramatically
to 1.4 by the end of the century, with an 80\% projection interval 1.0--2.0.
For context, in the U.S. this is currently 4.6
and is projected to decline to 1.8 by 2100.

This trend is well known for the U.S. and other developed countries
(\cite{Lee2011}) and features in political debate
and policy making there. What is perhaps surprising is that the same
trend is
projected for developing countries with young populations and currently high
potential support ratios like Egypt. Indeed, the decline is likely
to be even steeper in many developing countries than in developed
countries. Egypt in 2100 may well have an older population than the
U.S. or any other country in the world does now.
The projection intervals show that this overall trend
is essentially inevitable, even if there is some uncertainty about
the extent of the eventual decline.

\section{Accommodating Controversy About Ultimate Fertility Level via Bayes}
\label{sect-newtfr}
Our method for projecting the total fertility rate for all countries
was discussed during a three-day Expert Group Meeting convened by the UN
in December 2009 (\cite{EGM2009}) and was favorably assessed.
The predictive median from our method was then used as the
UN's (deterministic) projection of TFR in the WPP 2010 (\cite{UN2011}).
Apart from that, the UN used the same deterministic
projection method in WPP 2010 as in WPP 2008 and previous projections,
but is considering making future projections probabilistic.

Substantively, this led to projections of a slower decline in fertility
in Africa than had previously been expected. It also led to projections
of a slow increase in fertility in Europe, which had previously
been projected to remain at the sub-replacement level of 1.85 child per
woman once it reached this threshold. 
The previous UN projection
in WPP 2008 went up only to 2050 and projected a world population of 9
billion. The new WPP 2010 projection went up to 2100 for the first time
and projected a world population of 10 billion, a~billion more
(although for a longer time horizon).

Overall, the new projections were well received. However, there was
one critique, relating not to the statistical method, but to the
assumption\vadjust{\goodbreak}
in the model for TFR in Phase III
that asymptotically TFR oscillates around the approximate replacement
rate of 2.1, namely, that $\mu=2.1$ in (\ref{eq-PhaseIII}).

\citet{Basten2012} argued that the UN's assumption of an eventual recovery
of fertility toward replacement is not justified for five advanced
East Asian economies (Korea, Japan, Hong Kong, Singapore and Taiwan).
They pointed out that the national statistical agencies of these
countries project lower
fertility rates than does the UN, that the relevant scientific literature
does not suggest an increase in fertility in the short term,
that a recent unpublished survey of experts concluded that fertility
would not
increase as markedly as the UN predicts, and that current evidence about
fertility intentions does not suggest an immediate appetite for more
children in these countries. They also argued that the UN assumption
is based largely on European experience and that there is no reason to assume
that it will carry over to East Asia.
We interpret their arguments as implying that $\mu$ in (\ref{eq-PhaseIII})
may differ between East Asian countries and others, and that for
the five East Asian countries they consider, $\mu$ should be less than
the replacement level of 2.1.
\begin{figure*}

\includegraphics{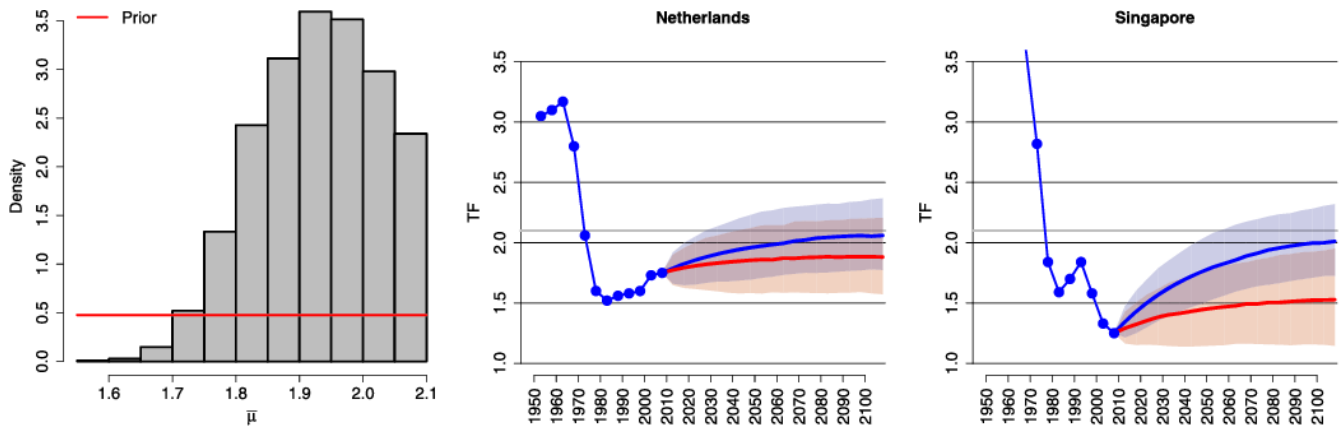}

\caption{Results for the Bayesian hierarchical model for Phase III fertility
(\protect\ref{eq-PhaseIIIhierlevel1})--(\protect\ref{eq-PhaseIIIhierprior}):
left: posterior distribution of $\bar{\mu}$, the world mean of the
country-specific asymptotes. Center and right: projections of the TFR
with 80\%
projection intervals for the Netherlands and Singapore under the
WPP 2010 model (\protect\ref{eq-PhaseIII}) in blue and the
Bayesian hierarchical model for Phase III in red.}
\label{fig-PhaseIIIhier}
\end{figure*}

We do not necessarily accept this critique, and there are counterarguments.
However, here we suggest a possible way to relax the assumptions
underlying the
Phase III fertility model (\ref{eq-PhaseIII}) so as to accommodate
the critique and make the model more fully data-based.
Instead of requiring every country in Phase III to follow the same model
(\ref{eq-PhaseIII}), we allow both $\mu$ and $\rho$ to vary between
countries following a Bayesian hierarchical model:
%
\begin{equation}\label{eq-PhaseIIIhierlevel1}
f_{c,t+1} - \mu_c = \rho_c (f_{c,t} -
\mu_c) + \varepsilon_{c,t},
\end{equation}
where $\varepsilon_{c,t} \stackrel{\mathrm{i.i.d.}}{\sim} N(0,\sigma
^2_{\varepsilon})$
and
%
\begin{equation}\label{eq-PhaseIIIhierlevel2}
\mu_c \sim TN_{[0, \infty)}\bigl(\bar{\mu}, \sigma^2_{\mu}
\bigr);\quad \rho_c \sim TN_{[0,1]}\bigl(\bar{\rho},
\sigma^2_{\rho}\bigr).\hspace*{-25pt}
\end{equation}

The prior distributions for the hyperparameters are as follows:
%
\begin{eqnarray}\label{eq-PhaseIIIhierprior}
\bar{\mu} &\sim& U[0, 2.1];\quad \sigma_{\mu} \sim U[0, 0.318];\nonumber\\
\bar{\rho} &\sim& U[0, 1];\quad \sigma_{\rho} \sim U[0, 0.289];\\
\sigma_{\varepsilon} &\sim& U[0, 0.5].\nonumber
\end{eqnarray}
The priors are chosen to be diffuse, except for the prior on $\bar{\mu}$,
the world mean of the country-specific asymptotes,
which is restricted to be no greater than the replacement level of 2.1.
Since the only critiques to date suggest that the UN's current choice
of 2.1 may be too high, this truncation seems to accommodate current
expert opinion.

Figure \ref{fig-PhaseIIIhier} shows some of the results of fitting the
Bayesian hierarchical model
(\ref{eq-PhaseIIIhierlevel1})--(\ref{eq-PhaseIIIhierprior})
to the data from the 21 countries that have entered Phase III.
The posterior distribution of $\bar{\mu}$, the world mean of the
country-specific TFR asymptotes, essentially excludes values below 1.6.
For most of the 21 countries, the projections are similar to those
from the WPP 2010 model (\ref{eq-PhaseIII}) of \citet{Alkema2011}
with fixed $\mu=2.1$;
this can be seen, for example, for the Netherlands in the middle panel of
Figure \ref{fig-PhaseIIIhier}.

The only one of Basten, Coleman and Gu's (\citeyear{Basten2012})
five advanced East Asian economies that has
entered Phase III is Singapore, for which the projections are shown in
the left
panel of Figure \ref{fig-PhaseIIIhier}. Remarkably, it is also the only
one of the 21 Phase III countries for which the projections differ
substantially between the \mbox{$\mu=2.1$} model and the Bayesian hierarchical
model with country-specific $\mu_c$.
The projection under the Bayesian hierarchical model
(\ref{eq-PhaseIIIhierlevel1})--(\ref{eq-PhaseIIIhierprior})
is much lower than under the previous model (\ref{eq-PhaseIII}),
as Basten et al. argued it should be. It asymptotes at 1.5 instead of 2.1.
This suggests that the data provide some support for Basten et al.'s contention,
and also that the proposed Bayesian hierarchical model for Phase III
can accommodate differences of this kind among countries.

\section{Discussion}
\label{sect-discussion}
We have developed a method for probabilistic population projections
for possible use by the United Nations in its biennial projections
of the populations of all countries. We have developed Bayesian hierarchical
models for projecting future overall levels of fertility and mortality
for each country, as measured by the total fertility rate and
life expectancy for females and males. Samples from the resulting
posterior predictive distributions are converted into age-specific
fertility and mortality rates. These are then used as inputs to the
cohort component projection method, yielding samples from the
posterior predictive distributions of any future population quantity
of interest. The method has performed reasonably well in terms of
out of sample predictive performance.

The method focuses on the overall levels of fertility and mortality,
traditionally viewed as the biggest sources of error in population
projections. However, there are other sources of uncertainty
not taken into account in our method. These include uncertainty about
the  base population, sex ratio at birth,  age structure of fertility and mortality and also about
net international migration, the latter being increasingly important
(\cite{NRC2000}). Not including these does not seem to have led to
substantial miscalibration of the method, but future work should
aim to incorporate these other sources of uncertainty.
Also, the method is not currently adapted to the countries with generalized
HIV/AIDS epidemics which comprise about 10\% of the world's population,
and it will be important to extend the method to these countries.

The UN projects populations not only for all countries, but also for
regions and other sets of countries including trading blocs and
economic, political and ecological groupings (\cite{UN2011aggregates}).
Aggregation of deterministic projections is straightforward: just add
up the
projections for the component countries. For probabilistic projections
it is not so simple, because between-country correlations have to be
taken into account. Our approach treats countries as exchangeable rather
than independent, but it does not model the additional correlation that
may exist between, for example, contiguous countries.
For life expectancy, work to date suggests that our model does account
for most of this correlation; there is little correlation between
forecast errors (\cite{RafteryChunn2012}).
For fertility there may be some excess between-country correlation,
and this needs to be accounted for in future work.

Our approach is largely Bayesian, and Bayesian thinking was essential
in overcoming the statistical challenges, including the limited amount
of data for each country and the differences and similarities among
countries in the way fertility and life expectancy have evolved.
The Bayesian approach allowed us to borrow strength from other countries
through the hierarchical model, thus avoiding instabilities in estimation.
It also gave us a way
to combine the posterior predictive distributions of fertility and mortality
with the cohort component model in a natural way, and to incorporate
external information about the asymptotic rate of increase in life
expectancy through its prior distribution.
In addition, it provided the basis for a way to accommodate
Basten, Coleman and Gu's (\citeyear{Basten2012}) critique of the UN's
assumption about the
equilibrium distribution of total fertility rate.
\citet{Fienberg2011} described several other major governmental
and policy problems for which Bayesian thinking proved useful.

We have adopted a fully Bayesian approach, however, only when it led to an
improved solution. For example, our model for the female-male gap in
life expectancy is not Bayesian, because there are enough data to estimate
the model reliably via maximum likelihood. Fully Bayesian estimation of
this model would give similar results and would be more complicated,
and so did not seem worth doing.

Several other Bayesian approaches to population projection have been proposed,
but these have focused mainly on projecting mortality.
\citet{GirosiKing08book} proposed a Bayesian method for forecasting
age-specific mortality that
can incorporate covariates. They showed that for short-term forecasts,
their method outperformed the
widely used time series method of \citet{LeeCarter1992} (without
covariates)
on average for 48 countries with better mortality data.
This result, obtained when covariates were used, requires additional data
that may not be reliable or even available in many countries.
The reliability of their approach remains unproven for medium- or long-term
projections due to the reliance on covariates and the difficulties
in predicting them beyond a few decades.
They did not give probabilistic forecasts,
although their method may in principle be
able to provide them.

\citet{Czadoetal2005} developed a Bayesian method for estimating
the Poisson log-bilin\-ear formulation of Brouhns, Denuit and Vermunt's
(\citeyear{Brouhns2002}) version of the Lee-Carter model.
Pedroza (\citeyear{Pedroza2006}) proposed a Bayesian approach to the Lee-Carter
model by accounting for the uncertainty in the age parameters as well
as the mortality index usually forecasted. While the latter two
approaches account for uncertainty in the Lee-Carter model, their
generalization to all countries is hindered by the nonavailability of
age-specific mortality rates. \citet{Daponte1997} developed a
Bayesian approach to the problem of reconstructing past populations,
which is different from the problem of projecting future populations
that we address here.

Expert-based probabilistic population projections have been produced by
Lutz and colleagues (Lutz, Sanderson and
Scherbov, \citeyear{Lutzetal98,Lutzetal04book,Lutzetal07projDB}).
However, this method is not explicitly based on available data,
and instead relies on a collection of experts and their ability
to specify probabilistic bounds, that may or may not be accurate
(\cite{Alho05book}). Thus, while like Bayesian approaches this method
uses expert knowledge, it does not update it formally using data, and
so is
not Bayesian in the usual sense.

Our model for the total fertility rate has been adopted by the UN
as the basis for its deterministic projections in WPP 2010 (\cite{UN2011}).
The UN also issued probabilistic projections using the methods
described here on an experimental basis in November 2012, at
\url{http://esa.un.org/unpd/ppp}.
The UN is considering issuing official probabilistic projections for
the first time in WPP 2014, using our methods.

The methods described here are implemented in the R packages
\texttt{bayesTFR} (\cite{Sevcikova2011}),
\texttt{bayesLife} (\cite{bayesLife}),
\texttt{bayesPop} (\cite{bayesPop})
and \texttt{bayesDem} (\cite{bayesDem}).

\section*{Acknowledgments}

This work was supported by Grants R01 HD054511 and R01 HD070936 from
the Eunice Kennedy Shriver National Institute of Child Health and Human
Development (NICHD) and a research grant from the National University
of Singapore. Its contents are solely the responsibility of the authors
and do not necessarily represent the official views of NICHD. Also, the
views expressed in this article are those of the authors and do not
necessarily reflect the views of the United Nations. Its contents have
not been formally edited and cleared by the United Nations. The
designations employed and the presentation of material in this article
do not imply the expression of any opinion whatsoever on the part of
the United Nations concerning the legal status of any country,
territory, city or area or of its authorities, or concerning the
delimitation of its frontiers or boundaries. The authors are grateful
to John Bongaarts, Thomas Buettner, Samuel Clark, Joel Cohen, Gerhard
Heilig, Ronald Lee, Nan Li, Sharon McGrayne, Hana
\v{S}ev\v{c}\'{\i}kov\'{a}, Hania Zlotnik, participants in the Bay Area
Colloquium in Population (BACPOP), the Editor, the Associate Editor and
two referees for very helpful comments and discussions.


%


\begin{thebibliography}{42}

\bibitem[\protect\citeauthoryear{Alho and Spencer}{2005}]{Alho05book}
%
\begin{bbook}[mr]
\bauthor{\bsnm{Alho},~\bfnm{Juha~M.}\binits{J.~M.}} \AND
\bauthor{\bsnm{Spencer},~\bfnm{Bruce~D.}\binits{B.~D.}}
(\byear{2005}).
\btitle{Statistical Demography and Forecasting}.
\bpublisher{Springer}, \blocation{New York}.
\bid{mr={2171856}}
\bptok{imsref}%
\end{bbook}
%
\endbibitem

\bibitem[\protect\citeauthoryear{Alkema et~al.}{2011}]{Alkema2011}
%
\begin{barticle}[auto:STB|2013/03/04|13:35:07]
\bauthor{\bsnm{Alkema},~\bfnm{L.}\binits{L.}},
\bauthor{\bsnm{Raftery},~\bfnm{A.~E.}\binits{A.~E.}},
\bauthor{\bsnm{Gerland},~\bfnm{P.}\binits{P.}},
\bauthor{\bsnm{Clark},~\bfnm{S.~J.}\binits{S.~J.}},
\bauthor{\bsnm{Pelletier},~\bfnm{F.}\binits{F.}},
\bauthor{\bsnm{Buettner},~\bfnm{T.}\binits{T.}} \AND
\bauthor{\bsnm{Heilig},~\bfnm{G.~K.}\binits{G.~K.}}
(\byear{2011}).
\btitle{Probabilistic projections of the total fertility rate for all
countries}.
\bjournal{Demography}
\bvolume{48}
\bpages{815--839}.
\bptok{imsref}%
\end{barticle}
%
\endbibitem

\bibitem[\protect\citeauthoryear{Basten, Coleman and Gu}{2012}]{Basten2012}
%
\begin{bmisc}[auto:STB|2013/03/04|13:35:07]
\bauthor{\bsnm{Basten},~\bfnm{S.~A.}\binits{S.~A.}},
\bauthor{\bsnm{Coleman},~\bfnm{D.~A.}\binits{D.~A.}} \AND
\bauthor{\bsnm{Gu},~\bfnm{B.}\binits{B.}}
(\byear{2012}).
\bhowpublished{Re-examining the fertility assumptions in the UN's 2010 World
Population Prospects: Intentions and fertility recovery in East Asia?
Presented at the Annual Meeting of the Population Association of
America, San
Francisco. Available at
\url{http://paa2012.princeton.edu/sessionViewer.aspx?SessionId=112}}.
\bptok{imsref}%
\end{bmisc}
%
\endbibitem

\bibitem[\protect\citeauthoryear{Brouhns, Denuit and
Vermunt}{2002}]{Brouhns2002}
%
\begin{barticle}[mr]
\bauthor{\bsnm{Brouhns},~\bfnm{Natacha}\binits{N.}},
\bauthor{\bsnm{Denuit},~\bfnm{Michel}\binits{M.}} \AND
\bauthor{\bsnm{Vermunt},~\bfnm{Jeroen~K.}\binits{J.~K.}}
(\byear{2002}).
\btitle{A~{P}oisson log-bilinear regression approach to the
construction of
projected lifetables}.
\bjournal{Insurance Math. Econom.}
\bvolume{31}
\bpages{373--393}.
\bid{doi={10.1016/S0167-6687(02)00185-3}, issn={0167-6687}, mr={1945540}}
\bptok{imsref}%
\end{barticle}
%
\endbibitem

\bibitem[\protect\citeauthoryear{Caswell}{2006}]{Caswell2006}
%
\begin{bbook}[auto:STB|2013/03/04|13:35:07]
\bauthor{\bsnm{Caswell},~\bfnm{H.}\binits{H.}}
(\byear{2006}).
\btitle{Matrix Population Models: Construction, Analysis and Interpretation}.
\bpublisher{Sinaurer}, \blocation{Sunderland}.
\bptok{imsref}%
\end{bbook}
%
\endbibitem

\bibitem[\protect\citeauthoryear{Czado, Delwarde and
Denuit}{2005}]{Czadoetal2005}
%
\begin{barticle}[mr]
\bauthor{\bsnm{Czado},~\bfnm{Claudia}\binits{C.}},
\bauthor{\bsnm{Delwarde},~\bfnm{Antoine}\binits{A.}} \AND
\bauthor{\bsnm{Denuit},~\bfnm{Michel}\binits{M.}}
(\byear{2005}).
\btitle{Bayesian {P}oisson log-bilinear mortality projections}.
\bjournal{Insurance Math. Econom.}
\bvolume{36}
\bpages{260--284}.
\bid{doi={10.1016/j.insmatheco.2005.01.001}, issn={0167-6687}, mr={2152844}}
\bptok{imsref}%
\end{barticle}
%
\endbibitem

\bibitem[\protect\citeauthoryear{Daponte, Kadane and
Wolfson}{1997}]{Daponte1997}
%
\begin{barticle}[auto:STB|2013/03/04|13:35:07]
\bauthor{\bsnm{Daponte},~\bfnm{B.~O.}\binits{B.~O.}},
\bauthor{\bsnm{Kadane},~\bfnm{J.~B.}\binits{J.~B.}} \AND
\bauthor{\bsnm{Wolfson},~\bfnm{L.~J.}\binits{L.~J.}}
(\byear{1997}).
\btitle{Bayesian demography: Projecting the Iraqi Kurdish population,
1977--1990}.
\bjournal{J. Amer. Statist. Assoc.}
\bvolume{92}
\bpages{1256--1267}.
\bptok{imsref}%
\end{barticle}
%
\endbibitem

\bibitem[\protect\citeauthoryear{El-Badry and Kono}{1986}]{ElBadryKono1986}
%
\begin{barticle}[auto:STB|2013/03/04|13:35:07]
\bauthor{\bsnm{El-Badry},~\bfnm{M.~A.}\binits{M.~A.}} \AND
\bauthor{\bsnm{Kono},~\bfnm{S.}\binits{S.}}
(\byear{1986}).
\btitle{Demographic estimates and projections}.
\bjournal{Population Bulletin of the United Nations}
\bvolume{19/20}
\bpages{35--43}.
\bptok{imsref}%
\end{barticle}
%
\endbibitem

\bibitem[\protect\citeauthoryear{Fienberg}{2011}]{Fienberg2011}
%
\begin{barticle}[mr]
\bauthor{\bsnm{Fienberg},~\bfnm{Stephen~E.}\binits{S.~E.}}
(\byear{2011}).
\btitle{Bayesian models and methods in public policy and government settings
(with discussion)}.
\bjournal{Statist. Sci.}
\bvolume{26}
\bpages{212--239}.
\bptok{imsref}%
\end{barticle}
%
\endbibitem

\bibitem[\protect\citeauthoryear{Girosi and King}{2008}]{GirosiKing08book}
%
\begin{bbook}[auto:STB|2013/03/04|13:35:07]
\bauthor{\bsnm{Girosi},~\bfnm{F.}\binits{F.}} \AND
\bauthor{\bsnm{King},~\bfnm{G.}\binits{G.}}
(\byear{2008}).
\btitle{Demographic Forecasting}.
\bpublisher{Princeton Univ. Press}, \blocation{Princeton, NJ}.
\bptok{imsref}%
\end{bbook}
%
\endbibitem

\bibitem[\protect\citeauthoryear{Hirschman}{1994}]{Hirschman1994}
%
\begin{barticle}[pbm]
\bauthor{\bsnm{Hirschman},~\bfnm{C.}\binits{C.}}
(\byear{1994}).
\btitle{Why fertility changes}.
\bjournal{Annu. Rev. Sociol.}
\bvolume{20}
\bpages{203--233}.
\bid{doi={10.1146/annurev.so.20.080194.001223}, issn={0360-0572},
pmid={12318868}}
\bptok{imsref}%
\end{barticle}
%
\endbibitem

\bibitem[\protect\citeauthoryear{Intergovernmental Panel on Climate
Change}{2007}]{IPCC2007}
%
\begin{bmisc}[auto:STB|2013/03/04|13:35:07]
\borganization{Intergovernmental Panel on Climate Change}
(\byear{2007}).
\bhowpublished{Climate Change 2007: Synthesis report, IPCC, Geneva,
Switzerland}.
\bptok{imsref}%
\end{bmisc}
%
\endbibitem

\bibitem[\protect\citeauthoryear{Keilman, Pham and Hetland}{2002}]{Keilman2002}
%
\begin{barticle}[auto:STB|2013/03/04|13:35:07]
\bauthor{\bsnm{Keilman},~\bfnm{N.}\binits{N.}},
\bauthor{\bsnm{Pham},~\bfnm{D.~Q.}\binits{D.~Q.}} \AND
\bauthor{\bsnm{Hetland},~\bfnm{A.}\binits{A.}}
(\byear{2002}).
\btitle{Why population forecasts should be probabilistic---illustrated
by the
case of Norway}.
\bjournal{Demographic Research}
\bvolume{6}
\bpages{409--454}.
\bptok{imsref}%
\end{barticle}
%
\endbibitem

\bibitem[\protect\citeauthoryear{Keyfitz and Caswell}{2005}]{KeyfitzCaswell2005}
%
\begin{bbook}[mr]
\bauthor{\bsnm{Keyfitz},~\bfnm{Nathan}\binits{N.}} \AND
\bauthor{\bsnm{Caswell},~\bfnm{H.}\binits{H.}}
(\byear{2005}).
\btitle{Applied Mathematical Demography},
\bedition{3rd} ed.
\bpublisher{Springer}, \blocation{New York}.
\bptok{imsref}%
\end{bbook}
%
\endbibitem

\bibitem[\protect\citeauthoryear{Lalic}{2011}]{Lalic2011}
%
\begin{bmisc}[auto:STB|2013/03/04|13:35:07]
\bauthor{\bsnm{Lalic},~\bfnm{N.}\binits{N.}}
(\byear{2011}).
\bhowpublished{Joint probabilistic projection of female and male life
expectancy.
Master's thesis, Dept. Statistics, Univ. Washington, Seattle, WA}.
\bptok{imsref}%
\end{bmisc}
%
\endbibitem

\bibitem[\protect\citeauthoryear{Lee}{2011}]{Lee2011}
%
\begin{barticle}[auto:STB|2013/03/04|13:35:07]
\bauthor{\bsnm{Lee},~\bfnm{R.~D.}\binits{R.~D.}}
(\byear{2011}).
\btitle{The outlook for population growth}.
\bjournal{Science}
\bvolume{333}
\bpages{569--573}.
\bptok{imsref}%
\end{barticle}
%
\endbibitem

\bibitem[\protect\citeauthoryear{Lee and Carter}{1992}]{LeeCarter1992}
%
\begin{barticle}[auto:STB|2013/03/04|13:35:07]
\bauthor{\bsnm{Lee},~\bfnm{R.~D.}\binits{R.~D.}} \AND
\bauthor{\bsnm{Carter},~\bfnm{L.}\binits{L.}}
(\byear{1992}).
\btitle{Modeling and forecasting the time series of US mortality}.
\bjournal{J. Amer. Statist. Assoc.}
\bvolume{87}
\bpages{659--671}.
\bptok{imsref}%
\end{barticle}
%
\endbibitem

\bibitem[\protect\citeauthoryear{Lee and
Tuljapurkar}{1994}]{LeeTuljapurkar1994}
%
\begin{barticle}[auto:STB|2013/03/04|13:35:07]
\bauthor{\bsnm{Lee},~\bfnm{R.~D.}\binits{R.~D.}} \AND
\bauthor{\bsnm{Tuljapurkar},~\bfnm{S.}\binits{S.}}
(\byear{1994}).
\btitle{Stochastic population forecasts for the {United States}: Beyond high,
medium, and low}.
\bjournal{J. Amer. Statist. Assoc.}
\bvolume{89}
\bpages{1175--1189}.
\bptok{imsref}%
\end{barticle}
%
\endbibitem

\bibitem[\protect\citeauthoryear{Leslie}{1945}]{Leslie1945}
%
\begin{barticle}[mr]
\bauthor{\bsnm{Leslie},~\bfnm{P.~H.}\binits{P.~H.}}
(\byear{1945}).
\btitle{On the use of matrices in certain population mathematics}.
\bjournal{Biometrika}
\bvolume{33}
\bpages{183--212}.
\bid{issn={0006-3444}, mr={0015760}}
\bptok{imsref}%
\end{barticle}
%
\endbibitem

\bibitem[\protect\citeauthoryear{Lutz and Samir}{2010}]{LutzSamir2010}
%
\begin{barticle}[auto:STB|2013/03/04|13:35:07]
\bauthor{\bsnm{Lutz},~\bfnm{W.}\binits{W.}} \AND
\bauthor{\bsnm{Samir},~\bfnm{K.~C.}\binits{K.~C.}}
(\byear{2010}).
\btitle{Dimensions of global population projections: What do we know about
future population trends and structures?}
\bjournal{Philosophical Transactions of the Royal Society B}
\bvolume{365}
\bpages{2779--2791}.
\bptok{imsref}%
\end{barticle}
%
\endbibitem

\bibitem[\protect\citeauthoryear{Lutz, Sanderson and
Scherbov}{1998}]{Lutzetal98}
%
\begin{barticle}[auto:STB|2013/03/04|13:35:07]
\bauthor{\bsnm{Lutz},~\bfnm{W.}\binits{W.}},
\bauthor{\bsnm{Sanderson},~\bfnm{W.~C.}\binits{W.~C.}} \AND
\bauthor{\bsnm{Scherbov},~\bfnm{S.}\binits{S.}}
(\byear{1998}).
\btitle{Expert-based probabilistic population projections}.
\bjournal{Population and Development Review}
\bvolume{24}
\bpages{139--155}.
\bptok{imsref}%
\end{barticle}
%
\endbibitem

\bibitem[\protect\citeauthoryear{Lutz, Sanderson and
Scherbov}{2004}]{Lutzetal04book}
%
\begin{bbook}[auto:STB|2013/03/04|13:35:07]
\bauthor{\bsnm{Lutz},~\bfnm{W.}\binits{W.}},
\bauthor{\bsnm{Sanderson},~\bfnm{W.~C.}\binits{W.~C.}} \AND
\bauthor{\bsnm{Scherbov},~\bfnm{S.}\binits{S.}}
(\byear{2004}).
\btitle{The End of World Population Growth in the 21st Century: New Challenges
for Human Capital Formation and Sustainable Development}.
\bpublisher{Sterling}, \blocation{Earthscan, VA}.
\bptok{imsref}%
\end{bbook}
%
\endbibitem

\bibitem[\protect\citeauthoryear{Lutz, Sanderson and
Scherbov}{2008}]{Lutzetal07projDB}
%
\begin{bmisc}[auto:STB|2013/03/04|13:35:07]
\bauthor{\bsnm{Lutz},~\bfnm{W.}\binits{W.}},
\bauthor{\bsnm{Sanderson},~\bfnm{W.~C.}\binits{W.~C.}} \AND
\bauthor{\bsnm{Scherbov},~\bfnm{S.}\binits{S.}}
(\byear{2008}).
\bhowpublished{IIASA's 2007 probabilistic world population projections, IIASA
world population program online data base of results. Available at
\url{http://www.iiasa.ac.at/Research/POP/proj07/index.html?sb=5}}.
\bptok{imsref}%
\end{bmisc}
%
\endbibitem

\bibitem[\protect\citeauthoryear{Myrskyla, Kohler and
Billari}{2009}]{Myrskyla2009}
%
\begin{barticle}[auto:STB|2013/03/04|13:35:07]
\bauthor{\bsnm{Myrskyla},~\bfnm{M.}\binits{M.}},
\bauthor{\bsnm{Kohler},~\bfnm{H.~P.}\binits{H.~P.}} \AND
\bauthor{\bsnm{Billari},~\bfnm{F.~C.}\binits{F.~C.}}
(\byear{2009}).
\btitle{Advances in development reverse fertility declines}.
\bjournal{Nature}
\bvolume{460}
\bpages{741--743}.
\bptok{imsref}%
\end{barticle}
%
\endbibitem

\bibitem[\protect\citeauthoryear{National Research Council}{2000}]{NRC2000}
%
\begin{bmisc}[auto:STB|2013/03/04|13:35:07]
\borganization{National Research Council}
(\byear{2000}).
\bhowpublished{\textit{Beyond Six Billion}: \textit{Forecasting the World's
Population}. National Academy Press, Washington, DC}.
\bptok{imsref}%
\end{bmisc}
%
\endbibitem

\bibitem[\protect\citeauthoryear{Oeppen and Vaupel}{2002}]{OeppenVaupel2002}
%
\begin{barticle}[auto:STB|2013/03/04|13:35:07]
\bauthor{\bsnm{Oeppen},~\bfnm{J.}\binits{J.}} \AND
\bauthor{\bsnm{Vaupel},~\bfnm{J.~W.}\binits{J.~W.}}
(\byear{2002}).
\btitle{Broken limits to life expectancy}.
\bjournal{Science}
\bvolume{296}
\bpages{1029--1031}.
\bptok{imsref}%
\end{barticle}
%
\endbibitem

\bibitem[\protect\citeauthoryear{Pedroza}{2006}]{Pedroza2006}
%
\begin{barticle}[pbm]
\bauthor{\bsnm{Pedroza},~\bfnm{Claudia}\binits{C.}}
(\byear{2006}).
\btitle{A Bayesian forecasting model: Predicting U.S. male mortality}.
\bjournal{Biostatistics}
\bvolume{7}
\bpages{530--550}.
\bid{doi={10.1093/biostatistics/kxj024}, issn={1465-4644}, pii={kxj024},
pmid={16484288}}
\bptok{imsref}%
\end{barticle}
%
\endbibitem

\bibitem[\protect\citeauthoryear{Preston, Heuveline and
Guillot}{2001}]{Preston2001}
%
\begin{bbook}[auto:STB|2013/03/04|13:35:07]
\bauthor{\bsnm{Preston},~\bfnm{S.~H.}\binits{S.~H.}},
\bauthor{\bsnm{Heuveline},~\bfnm{P.}\binits{P.}} \AND
\bauthor{\bsnm{Guillot},~\bfnm{M.}\binits{M.}}
(\byear{2001}).
\btitle{Demography: Measuring and Modeling Population Processes}.
\bpublisher{Blackwell}, \blocation{Malden, MA}.
\bptok{imsref}%
\end{bbook}
%
\endbibitem

\bibitem[\protect\citeauthoryear{Raftery et~al.}{2012}]{Raftery2012}
%
\begin{barticle}[auto:STB|2013/03/04|13:35:07]
\bauthor{\bsnm{Raftery},~\bfnm{A.~E.}\binits{A.~E.}},
\bauthor{\bsnm{Li},~\bfnm{N.}\binits{N.}},
\bauthor{\bsnm{{\v{S}}ev{\v{c}}{\'{\i}}kov{\'a}},~\bfnm{H.}\binits{H.}},
\bauthor{\bsnm{Gerland},~\bfnm{P.}\binits{P.}} \AND
\bauthor{\bsnm{Heilig},~\bfnm{G.~K.}\binits{G.~K.}}
(\byear{2012}).
\btitle{Bayesian probabilistic population projections for all countries}.
\bjournal{Proc. Natl. Acad. Sci. USA}
\bvolume{109}
\bpages{13915--13921}.
\bptok{imsref}%
\end{barticle}
%
\endbibitem

\bibitem[\protect\citeauthoryear{Raftery et~al.}{2013}]{RafteryChunn2012}
%
\begin{barticle}[auto:STB|2013/03/04|13:35:07]
\bauthor{\bsnm{Raftery},~\bfnm{A.~E.}\binits{A.~E.}},
\bauthor{\bsnm{Chunn},~\bfnm{J.~L.}\binits{J.~L.}},
\bauthor{\bsnm{Gerland},~\bfnm{P.}\binits{P.}} \AND
\bauthor{\bsnm{{\v{S}}ev{\v{c}}{\'{\i}}kov{\'a}},~\bfnm{H.}\binits{H.}}
(\byear{2013}).
\btitle{Bayesian probabilistic projections of life expectancy
for all
countries}.
\bjournal{Demography}
\bvolume{50}
\bpages{777--801}.
\bptok{imsref}%
\end{barticle}
%
\endbibitem

\bibitem[\protect\citeauthoryear{Seto, G{\"{u}}neral and
Hutyra}{2012}]{Seto2012}
%
\begin{barticle}[auto:STB|2013/03/04|13:35:07]
\bauthor{\bsnm{Seto},~\bfnm{K.~C.}\binits{K.~C.}},
\bauthor{\bsnm{G{\"{u}}neral},~\bfnm{B.}\binits{B.}} \AND
\bauthor{\bsnm{Hutyra},~\bfnm{L.~R.}\binits{L.~R.}}
(\byear{2012}).
\btitle{Global forecasts of urban expansion to 2030 and direct impacts on
biodiversity and carbon pools}.
\bjournal{Proc. Natl. Acad. Sci. USA}
\bvolume{109}
\bpages{16083--16088}.
\bptok{imsref}%
\end{barticle}
%
\endbibitem

\bibitem[\protect\citeauthoryear{United Nations}{2006}]{UN2006}
%
\begin{bmisc}[auto:STB|2013/03/04|13:35:07]
\borganization{United Nations}
(\byear{2006}).
\bhowpublished{World Population Prospects: The 2004 Revision. Volume III:
Analytical report, United Nations, New York}.
\bptok{imsref}%
\end{bmisc}
%
\endbibitem

\bibitem[\protect\citeauthoryear{United Nations}{2009}]{UN2009}
%
\begin{bmisc}[auto:STB|2013/03/04|13:35:07]
\borganization{United Nations}
(\byear{2009}).
\bhowpublished{World Population Prospects: The 2008 Revision. United Nations,
New York}.
\bptok{imsref}%
\end{bmisc}
%
\endbibitem

\bibitem[\protect\citeauthoryear{United Nations}{2011a}]{UN2011aggregates}
%
\begin{bmisc}[auto:STB|2013/03/04|13:35:07]
\borganization{United Nations}
(\byear{2011}a).
\bhowpublished{World Population Prospects: The 2010 Revision. Special
Aggregates---DVD-ROM Edition---Dataset in Excel and ASCII format. United
Nations Publication ST/ESA/SER.A/311, Population Division, Dept.
Economic and
Social Affairs, United Nations. Available at
\href{http://esa.un.org/unpd/wpp/Other-Information/WPP2010\_Special\%20Aggregates\%20-\%20list\%20of\%20groupings.pdf}
     {http://esa.un.org/unpd/wpp/Other-Information/}
     \href{http://esa.un.org/unpd/wpp/Other-Information/WPP2010\_Special\%20Aggregates\%20-\%20list\%20of\%20groupings.pdf}
     {WPP2010\_Special Aggregates - list of groupings.pdf}}.
\bptok{imsref}%
\end{bmisc}
%
\endbibitem

\bibitem[\protect\citeauthoryear{United Nations}{2011b}]{UN2011}
%
\begin{bmisc}[auto:STB|2013/03/04|13:35:07]
\borganization{United Nations}
(\byear{2011}b).
\bhowpublished{World Population Prospects: The 2010 Revision. United Nations,
New York}.
\bptok{imsref}%
\end{bmisc}
%
\endbibitem

\bibitem[\protect\citeauthoryear{United Nations Population
Division}{2009}]{EGM2009}
%
\begin{bmisc}[auto:STB|2013/03/04|13:35:07]
\borganization{United Nations Population Division}
(\byear{2009}).
\bhowpublished{Expert group meeting on recent and future trends in
fertility.
Available at
\url{http://www.un.org/esa/population/meetings/EGM-Fertility2009/egm-fertility2009.html}}.
\bptok{imsref}%
\end{bmisc}
%
\endbibitem

\bibitem[\protect\citeauthoryear{{\v{S}}ev{\v{c}}{\'{\i}}kov{\'
a}}{2011}]{baye%
sDem}
%
\begin{bmisc}[auto:STB|2013/03/04|13:35:07]
\bauthor{\bsnm{{\v{S}}ev{\v{c}}{\'{\i}}kov{\'a}},~\bfnm{H.}\binits{H.}}
(\byear{2011}).
\bhowpublished{bayesDem: Graphical user interface for bayesTFR,
bayesLife and
bayesPop. R package Version 1.6-0}.
\bptok{imsref}%
\end{bmisc}
%
\endbibitem

\bibitem[\protect\citeauthoryear{{\v{S}}ev{\v{c}}{\'{\i}}kov{\'a},
Alkema and
Raftery}{2011}]{Sevcikova2011}
%
\begin{bmisc}[auto:STB|2013/03/04|13:35:07]
\bauthor{\bsnm{{\v{S}}ev{\v{c}}{\'{\i}}kov{\'a}},~\bfnm{H.}\binits{H.}},
\bauthor{\bsnm{Alkema},~\bfnm{L.}\binits{L.}} \AND
\bauthor{\bsnm{Raftery},~\bfnm{A.~E.}\binits{A.~E.}}
(\byear{2011}).
\bhowpublished{bayesTFR: An R package for probabilistic projections of the
total fertility rate. \textit{Journal of Statistical Software} \textbf{43}
1--29}.
\bptok{imsref}%
\end{bmisc}
%
\endbibitem

\bibitem[\protect\citeauthoryear{{\v{S}}ev{\v{c}}{\'{\i}}kov{\'a} and
Raftery}{2011}]{bayesLife}
%
\begin{bmisc}[auto:STB|2013/03/04|13:35:07]
\bauthor{\bsnm{{\v{S}}ev{\v{c}}{\'{\i}}kov{\'a}},~\bfnm{H.}\binits{H.}}
\AND
\bauthor{\bsnm{Raftery},~\bfnm{A.~E.}\binits{A.~E.}}
(\byear{2011}).
\bhowpublished{bayesLife: Bayesian projection of life expectancy. R package
Version 0.4-0}.
\bptok{imsref}%
\end{bmisc}
%
\endbibitem

\bibitem[\protect\citeauthoryear{{\v{S}}ev{\v{c}}{\'{\i}}kov{\'a} and
Raftery}{2012}]{bayesPop}
%
\begin{bmisc}[auto:STB|2013/03/04|13:35:07]
\bauthor{\bsnm{{\v{S}}ev{\v{c}}{\'{\i}}kov{\'a}},~\bfnm{H.}\binits{H.}}
\AND
\bauthor{\bsnm{Raftery},~\bfnm{A.~E.}\binits{A.~E.}}
(\byear{2012}).
\bhowpublished{bayesPop: Probabilistic population projection. R package Version
1.0-3}.
\bptok{imsref}%
\end{bmisc}
%
\endbibitem

\bibitem[\protect\citeauthoryear{Whelpton}{1936}]{Whelpton1936}
%
\begin{barticle}[auto:STB|2013/03/04|13:35:07]
\bauthor{\bsnm{Whelpton},~\bfnm{P.~K.}\binits{P.~K.}}
(\byear{1936}).
\btitle{An empirical method for calculating future population}.
\bjournal{J. Amer. Statist. Assoc.}
\bvolume{31}
\bpages{457--473}.
\bptok{imsref}%
\end{barticle}
%
\endbibitem

\end{thebibliography}
\end{document}